\documentclass[12pt,preprint]{aastex}

\begin{document}

\title{Formation of ultra-compact blue dwarf galaxies and their evolution
into nucleated dwarfs}

\author{Kenji Bekki} 
\affil{
ICRAR,
M468,
The University of Western Australia
35 Stirling Highway, Crawley
Western Australia, 6009, Australia
}

\begin{abstract}
We propose that there is an evolutionary link between ultra-compact blue dwarf
galaxies (UCBDs) with active star formation
and nucleated dwarfs based on the results of numerical simulations
of dwarf-dwarf merging.
We consider the observational fact that low-mass dwarfs can be very gas-rich,
and thereby investigate the dynamical and chemical evolution of very gas-rich,
dissipative dwarf-dwarf mergers.
We find that the remnants of dwarf-dwarf mergers can be dominated by
new stellar populations formed from 
the triggered starbursts
and consequently can have blue colors and higher metallicities 
($Z\sim [0.2-1]Z_{\odot}$).
We also find that the remnants of these mergers can have rather high  mass-densities
($10^4 M_{\odot}$ pc$^{-3}$) within the central 10 pc
and small half-light radii ($40-100$ pc).
The radial stellar structures of some merger remnants are 
similar to those of nucleated dwarfs.
Star formation can continue in nuclear gas disks ($R<100$ pc) surrounding 
stellar galactic nuclei (SGNs) 
so that the SGNs can finally have  multiple stellar populations
with different ages and metallicities.
These very compact blue remnants can be identified as UCBDs soon after
merging and as nucleated dwarfs after fading of young stars.
We discuss these results in the context of the origins of metal-rich
ultra-compact
dwarfs (UCDs) and SGNs.
\end{abstract}


\section{Introduction}

Since the discovery of ultra-compact dwarf galaxies (``UCDs'') in
the Fornax clusters (Drinkwater et al. 2003),
the origin of UCDs has been extensively discussed both observationally
and theoretically (e.g., Bekki et al. 2003; 
Jones et al. 2006; Hilker et al. 2007; Gregg et al. 2009;
Brodie et al. 2011, B11; Mieske et al. 2013).
The typical half-light radii
of $r_{\rm h} \sim 20$ pc and luminosities of $L \sim 10^7 L_{\odot}$ 
are much larger than those of globular clusters 
(GCs; B11), which implies that the formation process of UCDs
is quite different from that of GCs.
C\^ote et al. (2006, C06) found that there is a similarity in color, luminosity
and size between UCDs and compact nuclei in a number of nucleated 
galaxies in the Virgo.
Recent observational studies of UCDs have revealed a number of  intriguing
physical properties of UCDs, such as a tight color-magnitude relation
for UCDs around M87 in the Virgo cluster (B11),
the presence of a supermassive black hole in an UCD around M60 (Seth et al. 2014),
and a young (1-2 Gyr old) 
stellar population in an UCD around NGC 4546 (Norris et al. 2015).

Although the formation of UCDs has been extensively discussed in the
context of transformation from nucleated dwarf and spiral galaxies
into naked stellar nuclei 
that can be identified as UCDs
(e.g., Bekki et al. 2001;  Pfeffer \& Baumgardt 2013),
the formation of nucleated galaxies itself has not been fully understood yet.
Both dissipationless merging of GCs
(Fellhauer \& Kroupa 2002)
and dissipative gas dynamics in the central regions of galaxies 
(e.g., Bekki 2007; Cole et al. 2014)
have been suggested to be important for the formation and growth of
stellar galactic nuclei (SGNs). 
Some UCDs and SGNs are observed to have high metallicities ($Z \sim Z_{\odot}$)
and young ages of less than 3 Gyr (e.g., Paudel et al. 2010; B11;
Norris et al. 2015), which appears to be very hard to be explained simply by
merging of metal-poor GCs and
by dissipative gas dynamics in  metal-poor dwarfs.

Recent observational studies of blue galaxies selected from
Sloan Digital Sky Survey (SDSS) have discovered 
nine ultracompact blue dwarf galaxies (``UCBDs'') with physical diameters
less than 1 kpc (Corbin et al. 2006). Such very compact galaxies were
also reported in previous observational studies of blue compact
dwarf galaxies (BCDs) by Kunth et al. (1988) and Doublier et al. (2000).
The central $B$-band surface brightness of 
POX 186, which is the prototype UCBD,  
is as high as those of SGNs (Doublier et al. 2000).
Furthermore, 
the central structures of young stellar population
in some UCBDs
appear to suggest that
recent collision/merging events 
has triggered the current active star formation of these
(Corbin et al. 2006).
However, it is unclear whether and how 
these UCBDs can possibly evolve into nucleated dwarfs
owing to the lack of theoretical work on this issue.

The purpose of this Letter is to propose that UCBDs can be formed
from merging of low-mass, gas-rich dwarfs based on 
the results of numerical simulations
on the physical properties of nuclear regions of dwarf-dwarf merger remnants.
Bekki (2008)
has suggested that galaxy merging can transform more massive 
gas-rich dwarf irregular
galaxies into blue compact dwarfs (BCDs) 
and further into nucleated dwarfs. However,
it did not investigate the dynamical and chemical properties
of their nuclear regions.
{\bf 
Stierwalt et al. (2015) have recently investigated
physical properties of interacting dwarfs
whereas Bekki (2015) has shown that low-mass gas-rich dwarf mergers ($10^8-10^9 M_{\odot}$)
can form extremely metal-poor galaxies.
However,  these authors did not discuss the central regions of 
merging dwarfs in detail.
Therefore, the present simulations 
are distinct from these recent ones on dwarf galaxy evolution.
}

\section{The model}

We perform numerical simulations of gas-rich dwarf-dwarf merging
and thereby investigate the dynamical and chemical properties
of the central regions of the merger remnants.
We  particularly investigate merging between
low-mass dwarf irregular (disk) galaxies 
with $10^9 {\rm M}_{\odot} \le  M_{\rm h}   \le 10^{10} M_{\odot}$,
where $M_{\rm h}$ is the total halo mass of a dwarf,
because we discuss the origins of UCBDs, UCDs, and SGNs in dwarfs.
In order to simulate the time evolution of
chemical abundances,  star formation rates (SFRs), and  gas contents,
we use our original chemodynamical simulation code 
with dust physics that can be run
on GPU machines (Bekki 2013).
A dwarf galaxy  consists of  dark matter halo,
stellar disk,  and  gaseous disk, and 
the total masses of dark matter halo, stellar disk, and  gas disk
are denoted as $M_{\rm h}$, $M_{\rm s}$, and $M_{\rm g}$, respectively.
The gas mass ratio ($M_{\rm g}/M_{\rm s}$) and baryonic mass fraction
($(M_{\rm s}+M_{\rm g})/M_{\rm h}$)
are denoted as $f_{\rm g}$, and $f_{\rm b}$
respectively,  for convenience.

The density distribution of the NFW
halo (Navarro, Frenk \& White 1996) suggested from CDM simulations
is adopted 
and the ``$c$-parameter''  ($c=r_{\rm vir}/r_{\rm s}$, where $r_{\rm vir}$
and $r_{\rm s}$ are  the virial
radius of a dark matter halo and the scale length of the halo)
and $r_{\rm vir}$ are chosen appropriately
for a given dark halo mass ($M_{\rm h}$)
by using the $c-M_{\rm h}$ relation
predicted by recent cosmological simulations (Neto et al. 2007).
The radial ($R$) and vertical ($Z$) density profiles of the stellar and gaseous disk are
assumed to be proportional to $\exp (-R/R_{0}) $ with scale
length $R_{0} = 0.4R_{\rm s}$  and to ${\rm sech}^2 (Z/Z_{0})$ with scale
length $Z_{0} = 0.02R_{\rm s}$, respectively.
Guided by observational results on $f_{\rm g}$ dependent
on $M_{\rm s}$ (Papastergis et al. 2012),
we investigate very gas-rich initial dwarf disks with $f_{\rm g}$ =1, 3, and 10
for $f_{\rm b}=0.006$, 0.018, and 0.06.
The initial gaseous metallicity is set to be $\log (Z/Z_{\odot}) = -1.6$.
The  Kennicutt-Schmidt law
(SFR$\propto \rho_{\rm g}^{\alpha_{\rm sf}}$;  Kennicutt 1998) 
with $\alpha_{\rm sf}=1.5$ 
and
the threshold gas density of  $\rho_{\rm th}=100$ atoms cm$^{-3}$ 
for star formation
are  adopted for modeling star formation in the present
study.

We mainly investigate major merging in which
two {\it equal-mass} dwarfs  merge with each other to form a new dwarf.
The initial distance
and the pericenter distance ($R_{\rm p}$)
of two interacting/merging dwarfs are set to be
$10R_{\rm s}$ and $0.5R_{\rm s}$, respectively.
The orbital eccentricity ($e_{\rm p}$) is
set to be 1 (i.e., parabolic encounter) for all models.
The spin of each galaxy in an  merging pair  is specified by two
angles $\theta_i$  (in units of degrees), where suffix $i$ is
used to identify each galaxy and $\theta_i$ is
the angle between the z axis and the vector of the angular momentum of a disk.
The orbital plane of a merging pair is set to be the same as the $x$-$y$
plane.
We show the results of the models with $\theta_1=30$ and $\theta_2=45$
in the present study.

Although we have investigated the time evolution
of dwarf-dwarf mergers for 1.1 Gyr in numerous models with different
$M_{\rm h}$, $R_{\rm s}$, $f_{\rm b}$, and $f_{\rm g}$, 
we show the results of 
four representative models (M1-M4) with 
$M_{\rm h}=3 \times 10^{9} M_{\odot}$
that show typical and important
behaviors of SGN and UCBD formation in the present study.
We will discuss the results of other models extensively 
in our forthcoming papers.
The fiducial model M1 has $M_{\rm h}=3 \times 10^9 M_{\odot}$,
$R_{\rm s}=298$pc, $f_{\rm b}=0.018$, and $f_{\rm g}=3$,
and the parameter values for other models are listed
in Table 1.
The total numbers of particles used for  a merger model
is $1.1 \times 10^6$, and the mass resolution for gaseous components
is $4.4 \times 10^3 {\rm M}_{\odot}$ in the fiducial model.
The gravitational softening length for stellar 
and gaseous components 
is set to be  6 pc for the fiducial model.
The softening length is different in models with different 
$M_{\rm h}$ and $R_{\rm s}$.

In order to estimate the $B$-band surface brightness ($\mu_{\rm B}$)
profiles of
merger remnants, 
we use the stellar population synthesis (SSP) code, ``MILES''
(Vazdekis et al. 2010).
Since the initial gaseous metallicity is low ($Z \sim 0.025Z_{\odot}$),
we adopt the SSP for $Z=0.05Z_{\odot}$ for a given age.
New stars formed from gas are assumed to have the $B$-band mass-to-light-ratios
($M/L$) of 0.09 that corresponds to a SSP with an age of 0.1 Gyr.
This is reasonable and realistic,
because we focus on young merger remnants dominated by young stars in the 
present study. Old stars are assumed to have $M/L=1.7$ that corresponds
to a SSP with an age of 5 Gyr. Based on the simulated mass profiles
of all stars and these $M/L$, we can estimate the $\mu_{\rm B}$ profiles
of merger remnants.
In the following,  $T$ in a simulation  represents the time that has elapsed since
the simulation started.

\section{Results}

Figure 1 shows the spatial distribution of old and new stars 
in the merger remnant at $T=1.1$ Gyr for the fiducial model (M1).
The remnant has a compact nucleus (SGN) composed almost exclusively  of new
stars so that it can be morphologically classified as a nucleated dwarf
elliptical or spheroidal with a very blue and bright nucleus
just after the merging.  This formation of the blue nucleus
is closely associated with the efficient inward  transfer of
a large amount of  gas caused by energy dissipation of shocked gas 
during merging.
As shown in Figure 2, the central surface mass density ($\Sigma$)
of new stars at $R=10$ pc can be as high as $10^4 M_{\odot}$,
which is more than an oder of magnitude higher than $\Sigma$ of old stars.
Feedback effects of SNe can not completely remove the remaining gas
from the nuclear region in this merger remnant so that
the gas density can be also high at $T=1.1$ Gyr.

As a result of dissipative formation of the very young SGN,
the central $B$-band surface brightness $\mu_{\rm B}$ can be as high as 
15 mag arcsec$^{-2}$, which is more than five magnitudes higher than
the initial $\mu_{\rm B}$ of old stars of merger progenitor dwarfs.
The contribution of new stars to $\mu_{\rm B}$ becomes dominant
at $R<200$ pc, which is reflected on the abrupt change in the slope of 
the $\mu_{\rm B}$ profile around $R=200$ pc. 
The half-light radius of the remnant is 44 pc, and $R_{25}$ ($R_{27}$)
where $\mu_{\rm B}$ becomes 25 (27) mag arcsec$^{-2}$,
is 426 (863) pc for this merger remnant. 
Thus, this young merger remnant can be identified as an UCBD owing to the very
small $r_{\rm h}$ and the low surface brightness of the outer stellar envelope
($\mu_{\rm B}>25$ mag arcsec$^{-2}$).

As shown in Figure 2, the radial $\Sigma$ and 
$\mu_{\rm B}$ profiles of merge remnants
depend on the initial mass densities, $f_{\rm g}$, and $f_{\rm b}$.
The less compact disk models, M2-M4, show lower $\Sigma$ of new stars
thus systematically lower $\mu_{\rm B}$ in comparison with the fiducial
model (M1). The radial $\mu_{\rm B}$ profiles for M2 and M3 are determined
almost exclusively by the spatial distribution of new stars for most parts of
the merger remnants. Therefore, the $\mu_{\rm B}$ profiles
are relatively smooth and do not look like the observed surface brightness
profiles of nucleated dwarfs 
(e.g., C06). On the other hand, the $\mu_{\rm B}$ profile
in M4 with smaller $f_{\rm g}$ is quite similar to  the observed ones
for nucleated dwarfs (e.g., C06).
If star formation in M4 is truncated by an external process  
(e.g., tidal or ram pressure stripping of gas), 
then this remnant can evolve into
a nucleated dwarf with lower central $\mu_{\rm B}$ 
within the next few Gyr after fading of the SGN.

Although the $\mu_{\rm B}$ profiles are quite different between
these four models, the central $\Sigma$ of gas can be rather high
in all of these models.
These high gaseous $\Sigma$ are due largely to the presence of
nuclear gas disks surrounding SGNs in merger remnants. 
As shown in Figure 3,  the gas disk sizes are about two times larger
than $r_{\rm h}$, and the disks have outer warp-like structures
(M1 and M2) and spiral
arms (M4).  Gas in the nuclear disks can be gradually 
transferred to the central regions of SGNs so that
new stars can form within SGNs slowly yet steadily after dwarf-dwarf merging.
These results strongly suggest that
nuclear gas disks are closely related to the origin of younger ages
observationally estimated for SGNs in some dwarfs and for some UCDs 
(e.g., B11).

{\bf 
The age-metallicity relation (AMR) in Figure 4 shows
that the merging gas-rich dwarfs can have higher 
metallicities in the central regions only in the later phases of merging.
The merging dwarfs have lower metallicities ($Z < 0.2Z_{\odot}$) in the early actively star-forming
phases when central mass concentration starts: 
dwarfs at these phases might be more similar to
UCBDs in terms  of metallicities.
They can finally have higher metallicities in the central regions after merging,
mainly because
chemical enrichment can proceed rapidly owing to the ejection of
a large amount of metals from SNe and AGB stars that are formed efficiently
during later starbursts in dwarf-dwarf merging. 
SN feedback effects can not completely suppress the formation of new stars in the central
regions of merging dwarfs in the present simulations. 
}

As a result of secondary starbursts,  SGNs of the merger remnants
can show large metallicity spreads of more than
0.5 dex among their stellar populations in M1-M4.
As shown in Figure 4, the simulated {\it internal} metallicity spread
is  more remarkable in M1 with a more compact initial gas disk,
because chemical enrichment proceeds longer and more efficiently in this model.
The final high metallicity ($Z \sim Z_{\odot}$) is consistent with
the observed $Z$ of some SGNs and UCDs, which implies that these 
high-metallicity objects could have been formed from merging
of gas-rich dwarf disks with higher stellar densities 
at a higher redshift.
The metallicities of SGNs in M2, M3, and M4 with lower initial stellar densities
are not so high ($Z<0.5 Z_{\odot}$), though they are higher than typical
metallicities of dwarfs.

The four models consistently show 
that younger stars are likely to be more metal-rich.
Although the present
simulations investigate only a relatively short term (1.1 Gyr) 
evolution of dwarf-dwarf merging,
it is highly likely that 
the age and metallicity spread among stars
of SGNs can be even larger in simulations for long-term  evolution.
SGNs can grow through gas accretion onto them and star formation
inside SGN, as long as the nuclear gas disks can exist.
The degrees of  internal age and metallicity spreads in each SGN, however,
would be very difficult
to be precisely estimated  in current spectroscopic  studies 
for nucleated dwarfs outside the Local
Group.

\section{Discussion and conclusions}

The present results provide the following four implications on the origins
of UCBDs, UCDs, and nucleated galaxies. 
First,  luminosity-weighted mean ages of stellar populations 
can be young in {\it some} UCDs.
If dwarf-dwarf merging occurs relatively recently,
the SGNs of the remnants can contain very young starbursts populations.
If the remnants subsequently lose their stellar envelopes through
``galaxy threshing'' (i.e., tidal stripping of outer stellar envelopes
of nucleated galaxies; Bekki et al 2001), then the naked nuclei
can be observed as young UCDs.
For example, if POX 186, which is the prototype UCBD, is stripped its
outer stellar envelope through galaxy threshing,
then it can be identified as a young UCD.
Indeed,  recent spectroscopic studies of UCDs have already found such UCDs
with young or intermediate-age stellar populations 
(e.g., Fig. 6 in B11). The present models also predict
that these apparently young UCDs have {\it internal}
metallicity spreads within their 
stellar populations.

Second, the observed very old ($\sim 13$ Gyr) and metal-rich 
($Z \sim Z_{\odot}$) UCDs (e.g., Fig. 6 in B11) 
could have been formed {\it directly} from  gas-rich dwarf-dwarf merging 
at a high redshift:
These old and metal-rich UCDs can not be explained simply by
the threshing scenario
(Bekki et al. 2003).
As shown in the present study,  the remnants of mergers with higher $f_{\rm g}$
are more likely to have higher central mass densities and more diffuse outer stellar
envelopes.
Therefore, if dwarf-dwarf merging with possibly very high $f_{\rm g}$ can occur at a very high redshift,
then the remnants can 
have very compact SGNs with high metallicities due to very efficient
chemical enrichment.
Such merger remnants could  be identified as UCDs even
without efficient galaxy threshing owing to the very low surface brightness of
the stellar envelopes.

Third, metal-poor  UCBDs with active star formation can finally become
more metal-rich nucleated dwarfs or UCDs with very faint stellar envelopes.
Although Corbin et al. (2006) showed that   UCBDs have 12+log(O/H)$<7.65$ 
($\sim 0.1 Z_{\odot}$), the present study suggests that 
at least UCBDs formed from merging can finally evolve into more metal-rich
nucleated  dwarfs (or UCDs)
owing to continuous chemical enrichment in the high-density gas disks.
Fourth,  SGNs can continue to grow by converting gas of the nuclear gas
disks around SGNs into new stars until the gas disks is destroyed or stripped
by some external processes (e.g., ram pressure stripping).
These nuclear gas disks might be responsible for the growth of 
massive black holes (MBHs) that coexist
with SGNs in low-mass galaxies.

It should be noted here that this merger-driven SGN formation in dwarfs
is only one of possible mechanisms of SGN formation, given that other mechanisms
have been already proposed (e.g., Bekki et al. 2006; Antonini et al. 2012).
Late-type spiral galaxies,
which should not have experienced any major merger events,
are observed to have nuclear star clusters or SGNs (e.g., B\"oker et al. 2002).
Furthermore, Lotz et al (2004) found that the SGNs of dwarf ellipticals in
the  Virgo and Fornax clusters have colors similar to (or slightly
redder than) those of GCs, which implies that metallicities of SGNs
might not be so different between SGNs and GCs.
Also,  only less than 10\% of BCDs can be classified as 
UCBDs with active  star formation  (Doublier et al. 2000).
These observations imply that
merger-driven metal-rich 
SGN formation might not be the major mechanism of SGN formation.

{\bf
Using cosmological simulations of subhalo interaction and merging in a cluster,
Knebe et al (2006) demonstrated that subhalo interaction is quite important
in the mass evolution of subhalos, because such interaction can tidally strip
their outer envelopes.
Cloet-Osselaer et al. (2014) have recently investigated merging histories of dwarfs with
$M_{\rm h} \ge 10^9 M_{\odot}$ by combining $N$-body/SPH simulations and the Press-Schechter
formalism and thereby found that dwarf-dwarf merging can trigger strong star-formation
episodes. Although these simulations clearly show the possibility of dwarf-dwarf interaction and merging,
the statistics on mass-ratios, orbital eccentricities, and pericenter distances
for dwarf mergers with $M_{\rm h}=10^9 
-10^{10} M_{\odot}$ in different environments
at different redshifts are yet to be extensively investigated
in a cosmological context.
}

If dwarf-dwarf merging is more frequent at higher redshifts ($z>1$),
then  young SGNs forming from such merging 
might be observed as unresolved point sources with $L\sim 10^7 L_{\odot}$
owing to the  very low surface brightness of the outer stellar envelopes.
The present study did not discuss some key properties of SGNs,
such as the small mass-ratios of SGNs
to their host galaxies (e.g., B\"oker et al. 2002; C06;
Georgiev et al. 2014),
the rotational kinematics of SGNs in some galaxies like NGC 4244 (Seth et al. 2008),
coexistence of massive black holes and SGNs in galaxies 
(e.g., Graham \& Spitler 2009). 
Various dynamical and hydrodynamical
processes such as dynamical interaction of MBH and SGNs 
(e.g., Bekki \& Graham 2010) might be important for
better understanding these properties of SGNs.
It is therefore our future study to investigate what physical processes
can be closely associated with the origin of  these properties.

\acknowledgments
I am   grateful to the referee, Brad Gibson,  for his constructive and
useful comments.

\begin{deluxetable}{lllll}
\footnotesize  
\tablecaption{ Description of the parameter values
for the representative four models.
\label{tbl-1}}
\tablewidth{-2pt}
\tablehead{
\colhead{  Model ID  \tablenotemark{a} } &
\colhead{  $f_{\rm b}$ \tablenotemark{b} } &
\colhead{  $f_{\rm g}$ \tablenotemark{c}} &
\colhead{  $R_{\rm s}$ (pc) \tablenotemark{d}} &
\colhead{  $r_{\rm h}$ (pc) \tablenotemark{e} } }
\startdata
M1 & 0.018 & 3.0 & 298 & 44 \\
M2 & 0.006 & 3.0 & 473 & 53  \\
M3 & 0.018 & 10.0 & 963 & 105  \\
M4 & 0.006 & 1.0 & 963 & 53 \\
\enddata
\tablenotetext{a}{
The initial total  mass of a dark matter halo  ($M_{\rm h}$) is set to be
$3 \times 10^9 M_{\odot}$ for all  models.
}
\tablenotetext{b}{
The initial baryonic mass fraction of a dwarf.
}
\tablenotetext{c}{
The initial gas mass ratio ($M_{\rm g}/M_{\rm s}$) of a dwarf.
}
\tablenotetext{d}{
The initial stellar disk size of a  dwarf.
}
\tablenotetext{e}{
The half-light radius (in the $B$-band)  of a merger remnant
at $T=1.1$ Gyr.
}
\end{deluxetable}

\begin{figure}
\plotone{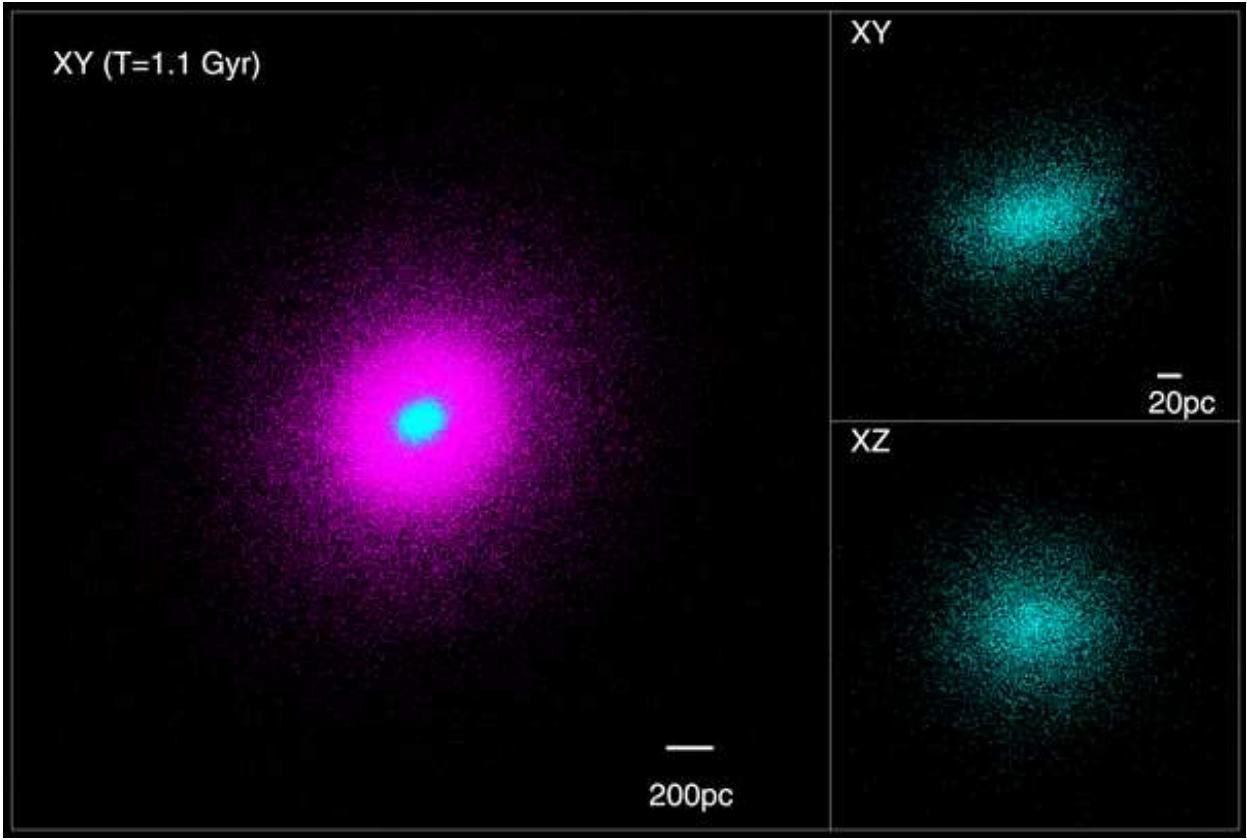}
\figcaption{
Distributions of old stars (magenta) and  new ones (cyan) 
of the dwarf-dwarf merger remnant projected onto
the $x$-$y$ plane  on a larger scale
(left) and  those projected onto the $x$-$y$ plane (upper right)
and onto the $x$-$z$ plane (lower right) on a smaller scale
in the fiducial model (M1)
at $T=1.1$ Gyr.
\label{fig-1}}
\end{figure}

\begin{figure}
\plotone{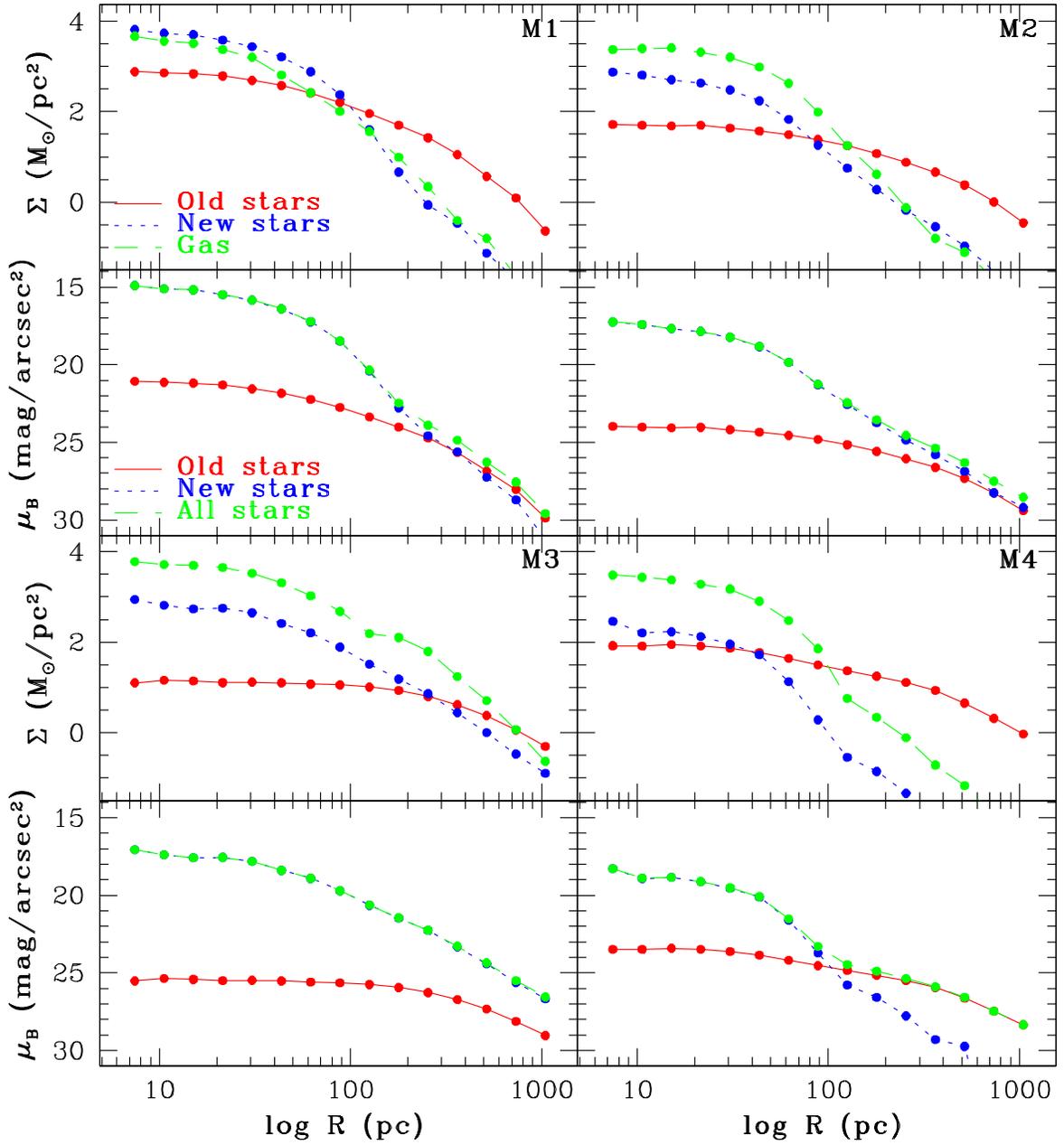}
\figcaption{
Radial surface mass density profiles ($\Sigma$) 
of old stars (red solid),  new stars (blue dotted),
and gas (green dashed) and the radial $B$-band surface brightness profiles
($\mu_{\rm B}$) for old stars (red solid), new stars (blue dotted),
and all stars (green dashed) of the merger remnant in M1-M4 models.
\label{fig-2}}
\end{figure}

\begin{figure}
\plotone{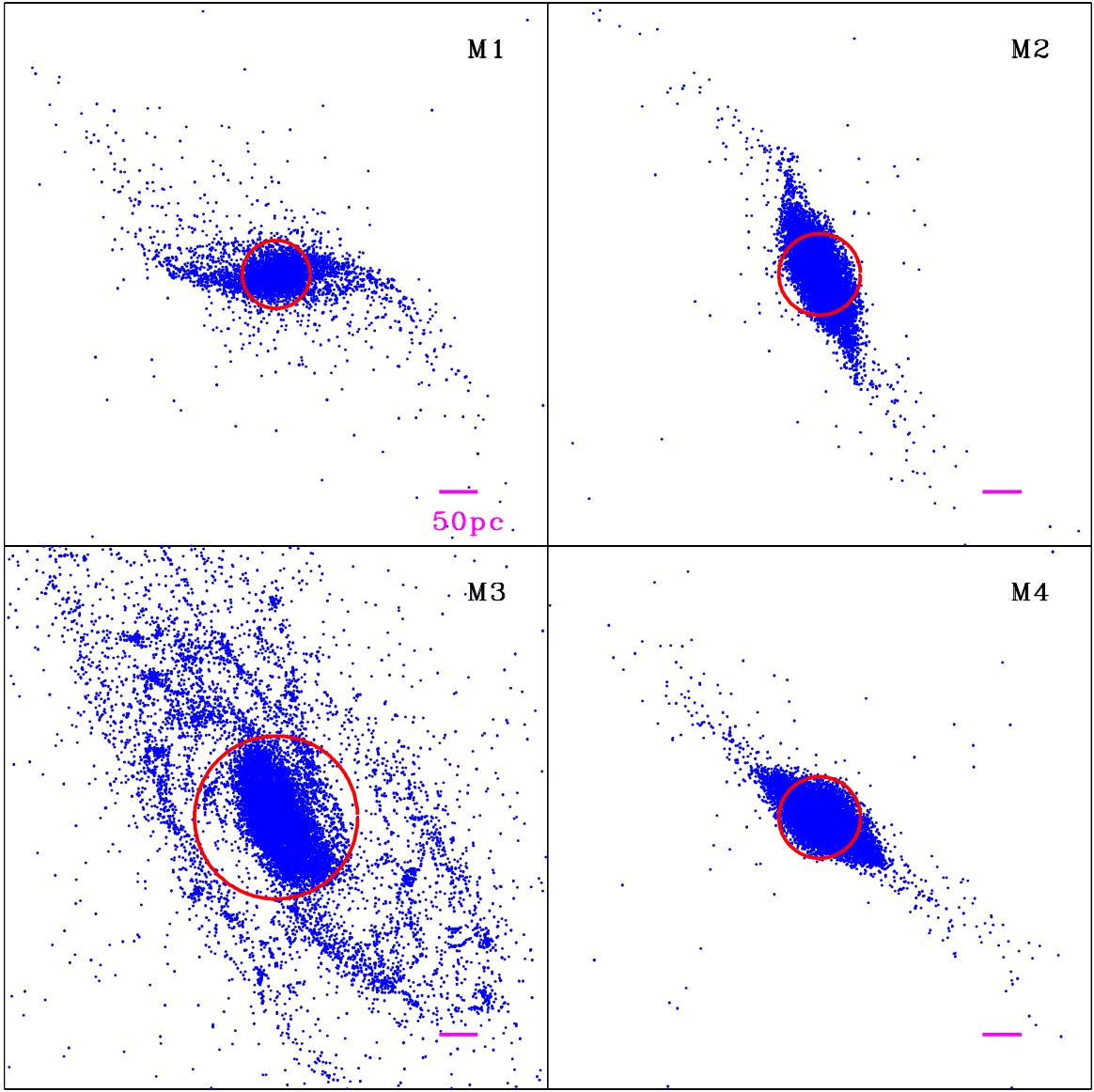}
\figcaption{
Distributions of gas around the newly formed SGNs in the remnants
of dwarf-dwarf mergers for the four models, M1-M4.
Each red circle indicates the half-light radius of each merger remnant.
\label{fig-3}}
\end{figure}

\begin{figure}
\plotone{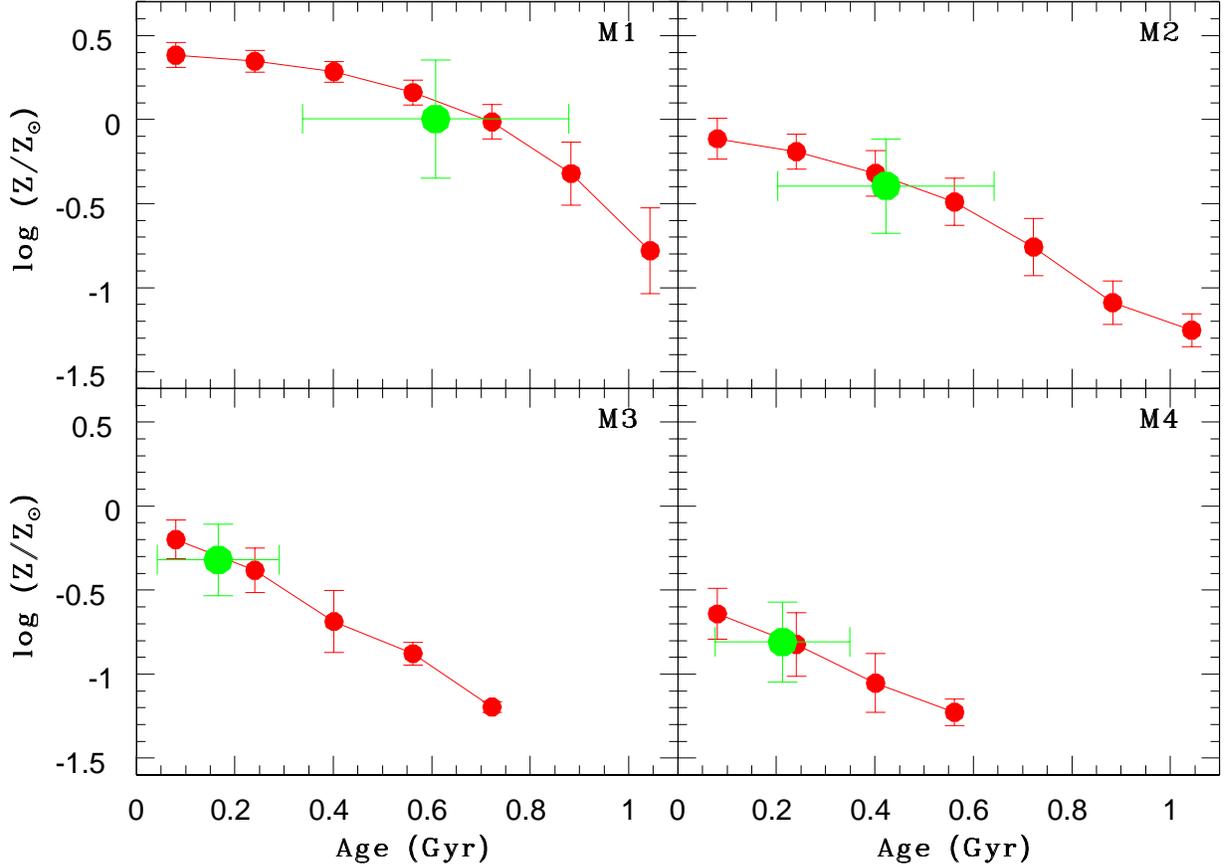}
\figcaption{
Age-metallicity relations (AMRs) of all new stars in SGNs 
of the merger remnants for the four models, M1-M4.
The red smaller circle indicates the mean metallicity at each age bin
whereas the green bigger circle indicates the mean metallicity and age
of the  new stars in each model.
The red error bar at each age bin 
shows the dispersion in  metallicities
of new stars at each age bin.
The dispersions of ages and metallicities for new stars of SGNs in each model
are indicated by green error bars.
The lack of data points for ages older than 0.6-0.7 Gyr in M3 and M4
is simply due to the lack of new stars formed before 0.6-0.7 Gyr ago.
\label{fig-4}}
\end{figure}

\end{document}